Snowmass 2013 White Paper
# Quartz Capillary Cladding Anthracene and Polycyclic Aromatic Hydrocarbon(PAH)-Core Scintillating/WLS Fibers for High Rates and Radiation Damage Resistance

**Fairfield University[1] - University of Iowa[2] - Argonne[3] Collaboration**

A. Albayrak-Yetkin[2], B. Bilki[2,3], J. Corso[2], G. Jennings[1], A. Mestvirisvilli[2], Y. Onel[2], I. Schmidt[2], C. Sanzeni[1], D. R. Winn[1]*, T. Yetkin[2]

*Contact/PI - winn@fairfield.edu
cell/txt/vm 1+203.984.3993

*Abstract:*
*Quartz capillary tube/fibers have been filled with anthracene by a melt and vacuum inbibition process to fabricate a scintillating core fiber. Other polcyclic aromatic hydrocarbons(PAH), such as p-Terphenyl (pTP), stilbene or naphthalene are also well-suited to scintillating/shifting fiber cores. The resulting scintillating core with quartz cladding capillary fibers (250-750 µm cores) had a high specific light output when tested with muons (8 p.e. per MIP). These PAH core quartz capillary cladding scintillating/shifting optical fibers have the potential of high radiation resistance, fast response, and are applicable to many energy and intensity frontier experiments.*

*Introduction:* Scintillating or wavelength-shifting (WLS) fibers[1] suitable for fast and rad-hard detectors were constructed with the following precepts:

(1) PAH (polycyclic aromatic hydrocarbon) organic scintillators/WLS, such as anthracene or pTerphenyl (pTP) are rad-hard because of the strong bonds between benzene-rings, and/or the scintillation mechanism which de-excites a pair of excited molecules via vibrations[2,3] (see the cartoon in Fig 1 for anthracene, as an example, and similar for the other PAH). The soft material in the solidified state is easily thermally annealed, and not amenable to permanent color centers. Similarly, p-Terphenyl(pTP), and other polycyclic aromatic hydrocarbons(PAH), consisting largely of linked benzene rings, are radiation-resistant due to rapid self-annealing[4,5] at doses up to 100's of MRad[6]. Deposition by various forms of PVD, especially simple thermal evaporation, result in useful films deposited on amorphous substrates (glass, quartz)[7].

(2) Many PAH have decay constants ~25 ns or less, matched to LHC (and beyond) interaction rates. These include pTerphenyl, stilbene, naphthalene and others (Fig.1).

(3) The peak light emission wavelengths can be ~380-520 nm, good for long transmission & PMT quantum efficiencies (Fig.1).

(4) The index of refraction is typically ~1.6, good for total internal reflection waveguiding with quartz (n=1.52) capillaries or tubes as a cladding (quartz is rad-hard). NA = 0.43 for Anthracene/Quartz clad and the light trapping is >3% of generated light.

(5) The melting points are modest, between 80°C-220°C, but must be done under vacuum or inert atmosphere to prevent oxidation – in the present work we use a partial vacuum.

(6) The confining and gas-impervious quartz-capillary:
 a) forces a liquid core as it solidifies to a continuous amorphous but nearly ordered solid fiber;
 b) keeps the Gibbs free energy negative, for the reverse reaction to occur after any radiolysis; and
 c) prevents interactions with $O_2$.

These properties allow amorphous or microcrystalline soft-material based core fibers to be used for calorimetry or preradiators in high radiation environments.
The meltable scintillating/shifting core materials include:

---

*Anthracene:*

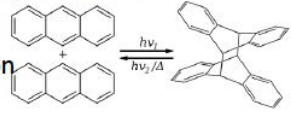

### Anthracene Properties

- Anthracene
  - Vibrational energy dissipation
    - Reduces raddam
  - Melts: 218° C  Boils: 340° C; soluble methanol
  - Densities: Solid 1.25 g/cm³  Liquid 0.97 g/cm³
  - Index: 1.595
  - Bright scintillator x2.3 of NaI;
  - emission between 350-500nm
  - Decay time: <30ns

*Stilbene(trans):*   70% anthracene; $\tau_{decay}$~3.5ns; n=1.64; λ~375-450 nm; ρ=1.22; $T_m$=124°C 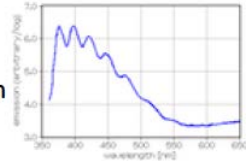

*pTerphenyl(pTP):* 120% anthracene; $\tau_{decay}$~3.0 ns; n=1.64, λ~375-450 nm; ρ=1.22; $T_m$=214°C 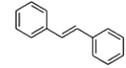

*Napthalene*:   ~50% anthracene; $\tau_{decay}$~5  ns; n=1.58; λ~330-350 nm;  ρ=1.14; $T_m$=  80°C 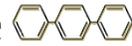

---

**Figure 1: Properties of polycyclic aromatic hydrocarbons (PAH) suitable for quartz capillary scintillating cores.**

**Prototypes:** Quartz capillaries were obtained from Polymicro[8], with core diameters between 250-750 μm, with an external low index doped quartz cladding. A vacuum tube furnace was assembled with temperature control as shown in Fig. 2 below. An inner pyrex vacuum burette with re-entrant parts for anthracene melt was used to hold the capillaries, with one end resting in anthracene powder. The partial vacuum was ~2-3 x $10^{-3}$ Bar. We recommend better vacuum, and prefilling with nitrogen or Argon. After a temperature threshold was crossed, typically about 5-8°C past the melting point, the liquid anthracene uniformly filled the fibers, even energetically fountaining from the top, presumably due to capillary action. More elevated temperatures caused the anthracene to evacuate the fibers, or to begin to oxidize. The temperature was brought down over ~30 minutes. Optimization of the process requires further work. Examination of the resulting fibers with a microscope showed uniform cores with essentially no inclusions, bubbles, or defects in most fibers.

**Test Results:** Strong scintillation was observed with a UV light on one end, with the scintillation light shining from the far ends (Figs. 4). A $^{137}$Cs source also evidenced scintillation detected by a PMT. The fibers were then mounted as a close-packed 2 layer ribbon in an Al holder, and read on each end by an 8mm diameter Hamamatsu metal envelope PMT. They were exposed to a muon beam in the CERN H2 test beam, as defined by a wire hodoscope and muon trigger. The ADC spectra are shown in the Figures below. Remarkably, ~9-10 p.e. per end was measured as shown in Figs. 5,6; this rivals or exceeds existing plastic scintillating fibers, and was consistent with estimates (see the figures). The fibers were too short (~25cm) to observe attenuation effects.

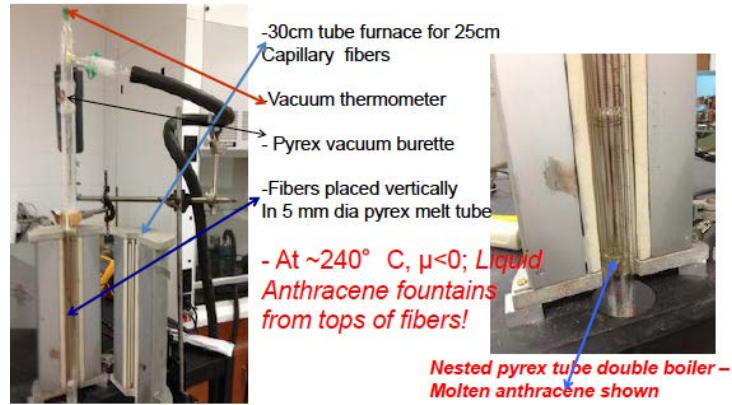

***Fig. 2:*** The vacuum tube furnace used to fill quartz capillaries with anthracene cores. The capillary action at ~240°C fountains the molten anthracene out from the tops of the quartz capillaries, and then a tapered cooling cycle forms perfect scintillating core fibers in the capillaries.

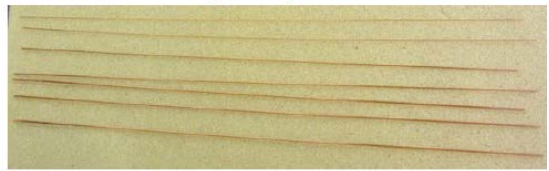

***Fig. 3:*** Fabricated anthracene core fibers with quartz capillary claddings.

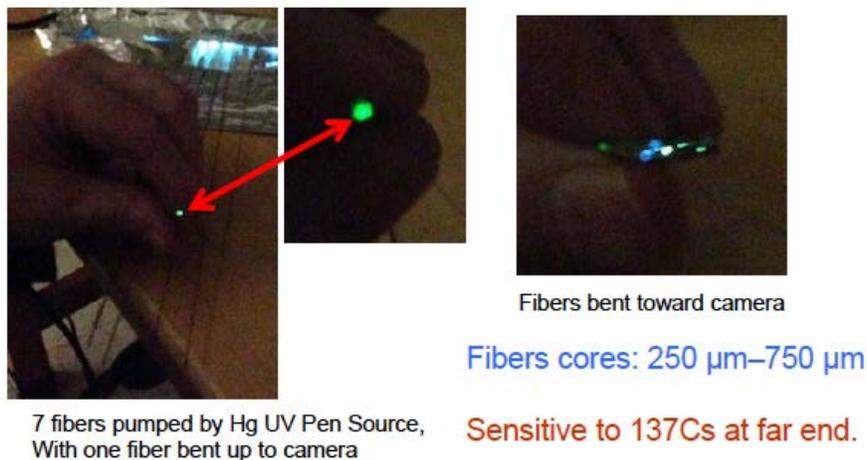

***Fig. 4:*** Preliminary tests of the WLS and scintillation properties of the anthracene core quartz capillary optical fibers, using a UV light source to pump one end of the fibers, the photographs show the far end scintillation. $^{137}$Cs sources also showed scintillation, but were too weak to photograph; a PMT was integrated with a fA meter. The fibers were inspected with a microscope for obvious core defects.

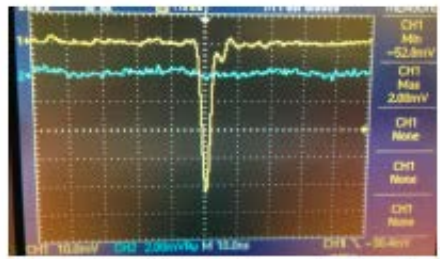

**Expected Anthracene Fiber Pulse:**
~200 KeV/mm x 0.25mm x 40 photons/KeV
x 2% transmission x 20% QE ~ 8 p.e.
**Typical Observed Pulse:**
~ 8-9 p.e.

Typical pulse in 80GeV e⁻ beam

*Fig.5:* Typical pulses from one end of the anthracene fibers exposed to MIPS. Pulse width 10%-10% <10 ns.

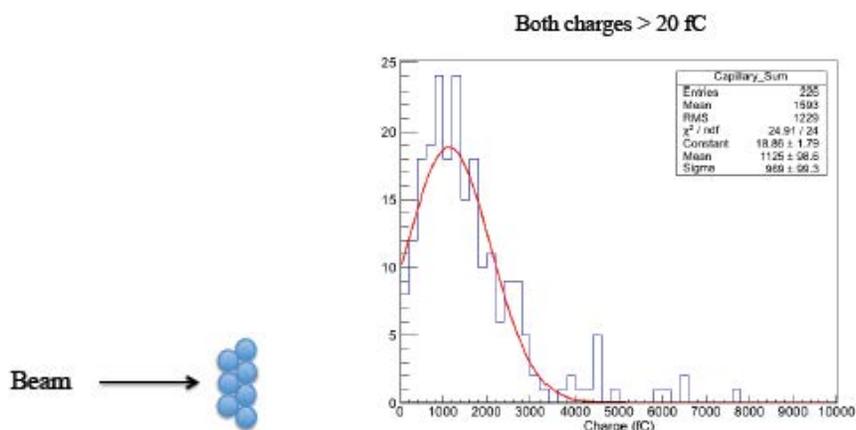

*Fig. 6:* The electron beam of 2x2 cm was perpendicular to the fiber 2 layer ribbon (fibers as in Fig. 3), which were arranged as shown at left, and read out from both fiber ends with 8mm diameter Hamamatsu PMT. The RH figure shows the ADC distribution from the sum of both fiber ends, which corresponds to a mean of about 18-20 p.e. or ~9-10p.e. per end. (The threshold cut of 20 fC per end corresponds to ~¼ p.e. per end.)

**Summary: Anthracene Core Quartz-Cladding Fibers:**
These novel fibers show great promise, with a high light output, exceeding many existing fibers. Future work is planned to include radiation damage studies, the fabrication of longer fibers, and attenuation length measurements. We request support for continuing this development.